\documentclass[12pt]{article}
\usepackage{epsfig}
\newcommand{\ud}{\mathrm{d}}
\begin{document}
\title{Variational--Wavelet Approach to RMS Envelope Equations}
\author{ANTONINA N. FEDOROVA and  MICHAEL G. ZEITLIN\\
Institute of Problems of Mechanical Engineering,\\
 Russian Academy of Sciences, 199178, Russia, St.~Petersburg,\\
  V.O., Bolshoj pr., 61, \\E-mail: zeitlin@math.ipme.ru, \\
http://www.ipme.ru/zeitlin.html\\
http://www.ipme.nw.ru/zeitlin.html}
\date{February 21, 2000}
\maketitle
\thispagestyle{empty}
\abstract{
We
present applications of variational--wavelet approach to nonlinear (rational)
rms envelope equations.
 We have the solution as
a multiresolution (multiscales) expansion in the base of compactly
supported wavelet basis.
We give extension of our  results to the cases of periodic beam
motion and arbitrary variable coefficients.
Also we consider more flexible variational
method which is based on biorthogonal wavelet approach.
}
\vspace*{30mm}

\begin{center}
Paper presented at:\\
Second ICFA Advanced Accelerator Workshop\\
THE PHYSYCS OF HIGH BRIGHTNESS BEAMS\\
UCLA Faculty Center, Los Angeles\\
November 9-12, 1999

\end{center}
\newpage
\section{Introduction}

In this paper we consider the applications of a new numerical-analytical 
technique which is based on the methods of local nonlinear Fourier
analysis or Wavelet analysis to the nonlinear beam/accelerator physics
problems related to root-mean-square (rms) envelope dynamics [1].
Such approach may be useful in all models in which  it is 
possible and reasonable to reduce all complicated problems related with 
statistical distributions to the problems described 
by systems of nonlinear ordinary/partial differential 
equations. In this paper we  consider  approach based on 
the second moments of the distribution functions for  the calculation
of evolution of rms envelope of a beam.

The rms envelope equations are the most useful for analysis of the 
beam self--forces (space--charge) effects and also 
allow to consider  both transverse and longitudinal dynamics of
space-charge-dominated relativistic high--bright\- ness
axisymmetric/asymmetric beams, which under short laser pulse--driven
radio-frequency photoinjectors have fast transition from nonrelativistic
to relativistic regime [2]-[3].
   
From the formal point of view we may consider rms envelope equations 
after straightforward transformations to standard Cauchy form 
as a system of nonlinear differential equations 
which are not more than rational (in dynamical variables). 
Such rational type of
nonlinearities allow us to consider some extension of results from 
[4]-[12], which are based on application of wavelet analysis technique to 
variational formulation of initial nonlinear problem.  

 Wavelet analysis is a relatively novel set of mathematical
methods, which gives us a possibility to work with well-localized bases in
functional spaces and give for the general type of operators (differential,
integral, pseudodifferential) in such bases the maximum sparse forms.

An example of such type of basis is demonstrated on Fig.~1.

Our approach in this paper is based on the generalization [13] of variatio\-nal\--wavelet 
approach from [4]-[12],
which allows us to consider not only polynomial but rational type of 
nonlinearities. 

So, our variational-multiresolution approach gives us 
possibility to construct explicit numerical-analytical
solution for the following systems of nonlinear differential 
equations 
\begin{equation}
\dot{z}=R(z,t) \quad \mbox{or} \quad Q(z,t)\dot{z}=P(z,t),
\end{equation}
where $z(t)=(z_1(t),...,z_n(t))$ is the vector of dynamical variables 
$z_i(t)$,

$R(z,t)$ is not more than rational function of z,

$P(z,t), Q(z,t)$ are not more than polynomial functions of z and 
P,Q,R have arbitrary dependence of time. 

The solution has the following form
\begin{equation}\label{eq:z}
z(t)=z_N^{slow}(t)+\sum_{j\geq N}z_j(\omega_jt), \quad \omega_j\sim 2^j
\end{equation}
which corresponds to the full multiresolution expansion in all time 
scales.
Formula (\ref{eq:z}) gives us expansion into a slow part $z_N^{slow}$
and fast oscillating parts for arbitrary N. So, we may move
from coarse scales of resolution to the 
finest one for obtaining more detailed information about our dynamical process.
The first term in the RHS of equation (2) corresponds on the global level
of function space decomposition to  resolution space and the second one
to detail space. In this way we give contribution to our full solution
from each scale of resolution or each time scale (detailed description
we give in part 3.2 and numerical illustration in part 7 below).
The same is correct for the contribution to power spectral density
(energy spectrum): we can take into account contributions from each
level/scale of resolution.

In part 2 we describe the different forms of rms equations.
Starting  in part 3.1 from  variational formulation of
initial dynamical problem we
construct via multiresolution analysis (3.2)
 explicit representation for all dynamical variables in the base of
compactly supported (Daubechies) wavelets. Our solutions (3.3)
are parametrized
by solutions of a number of reduced algebraical problems one from which
is nonlinear with the same degree of nonlinearity and the rest  are
the linear problems which correspond to particular
method of calculation of scalar products of functions from wavelet bases
and their derivatives. Then  we consider further extension of our
previous results.  In part 4 we consider modification of our
construction to the periodic case,
in part 5 we consider generalization of our approach
 to variational formulation in the biorthogonal bases of compactly
supported wavelets and in part 6 to the case of variable coefficients.
In part 7 we consider results of numerical calculations.

\begin{figure}[ht]
\centering
\epsfig{file=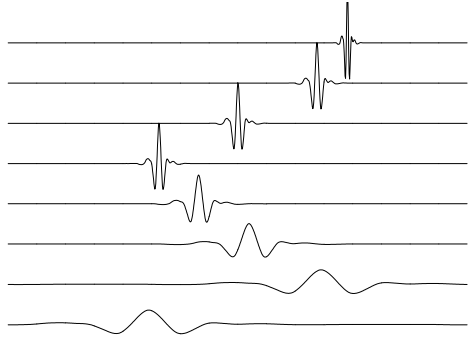, width=82mm, bb=0 200 599 590, clip}
\caption{Wavelets at different scales and locations.}
\end{figure}

\section{RMS Equations}
Below we consider a number of different forms of RMS envelope equations,
which are from the formal  point of view
not more than nonlinear differential equations with rational
nonlinearities and variable coefficients.
Let $f(x_1,x_2)$ be the distribution function which gives full information about 
noninteracting ensemble of beam particles regarding to trace space or 
transverse phase coordinates $(x_1,x_2)$. Then (n,m) moments are:
\begin{equation}
\int\int x_1^n x_2^m f(x_1,x_2)\ud x_1\ud x_2
\end{equation}
The (0,0) moment gives normalization condition on the distribution.
The (1,0) and (0,1) moments vanish when a beam is aligned to its axis.
Then we may extract the first nontrivial bit of `dynamical information' from 
the second moments
\begin{eqnarray}
\sigma_{x_1}^2&=&<x_1^2>=\int\int x_1^2 f(x_1,x_2)\ud x_1\ud x_2 \nonumber\\
\sigma_{x_2}^2&=&<x_2^2>=\int\int x_2^2 f(x_1,x_2)\ud x_1\ud x_2 \\
\sigma_{x_1 x_2}^2&=&<x_1 x_2>=\int\int x_1 x_2 f(x_1,x_2)\ud x_1\ud x_2 \nonumber
\end{eqnarray}
RMS emittance ellipse is given by
\begin{equation}
\varepsilon^2_{x,rms}=<x_1^2><x_2^2>-<x_1 x_2>^2
\end{equation}
 Expressions for twiss  parameters are also based on the second moments.

We will consider the following particular
cases of rms envelope equations, which described evolution
of the moments (4) ([1]-[3] for full designation):

for asymmetric beams we have the system of two envelope equations
of the second order for $\sigma_{x_1}$ and $\sigma_{x_2}$:

\begin{eqnarray}
\sigma^{''}_{x_1}+\sigma^{'}_{x_1}\frac{\gamma '}{\gamma}+
\Omega^2_{x_1}\left(\frac{\gamma '}{\gamma}\right)^2\sigma_{x_1}&=&
\frac{I}{I_0(\sigma_{x_1}+\sigma_{x_2})\gamma^3}+
\frac{\varepsilon^2_{nx_1}}{\sigma_{x_1}^3\gamma^2},\nonumber\\
\sigma^{''}_{x_2}+\sigma^{'}_{x_2}\frac{\gamma '}{\gamma}+
\Omega^2_{x_2}\left(\frac{\gamma '}{\gamma}\right)^2\sigma_{x_2}&=&
\frac{I}{I_0(\sigma_{x_1}+\sigma_{x_2})\gamma^3}+
\frac{\varepsilon^2_{nx_2}}{\sigma_{x_2}^3\gamma^2}
\end{eqnarray}

the envelope equation for an axisymmetric beam is

\begin{equation}
\sigma^{''}+\sigma^{'}\frac{\gamma '}{\gamma}+
\Omega^2\left(\frac{\gamma '}{\gamma}\right)^2\sigma=
\frac{k_s}{\sigma\gamma^3}+
\frac{\varepsilon^2_{n,th}}{\sigma^3\gamma^2}
\end{equation}

Also we have related Lawson's equation for evolution of the rms
envelope in the paraxial limit, which governs evolution of cylindrical
symmetric envelope under external linear focusing channel
of strenghts $K_r$:

\begin{equation}
\sigma^{''}+\sigma^{'}\left(\frac{\gamma '}{\beta^2\gamma}\right)+
K_r\sigma=\frac{k_s}{\sigma\beta^3\gamma^3}+
\frac{\varepsilon^2_n}{\sigma^3\beta^2\gamma^2},\nonumber
\end{equation}

where

\begin{equation}
K_r\equiv -F_r/r\beta^2\gamma mc^2, \ \ \
 \beta\equiv \nu_b/c=\sqrt{1-\gamma^{-2}}\nonumber
\end{equation}

After transformations to Cauchy form we can see that
all this equations from the formal point of view are not more than
ordinary differential equations with rational nonlinearities
and variable coefficients and correspond to the form (1)
(also,we may consider regimes in which $\gamma$, $\gamma'$
are not fixed functions/constants but satisfy some additional differential constraint/equation,
but this case does not change our general approach).

\section{Rational Dynamics}

The first main part of our consideration is some variational approach
to this problem, which reduces initial problem to the problem of
solution of functional equations at the first stage and some
algebraical problems at the second stage.
We have the solution in a compactly
supported wavelet basis.
Multiresolution expansion is the second main part of our construction.
The solution is parameterized by solutions of two reduced algebraical
problems, one is nonlinear and the second are some linear
problems, which are obtained from one of the next wavelet
constructions:  the method of Connection
Coefficients (CC), Stationary Subdivision Schemes (SSS).

\subsection{ Variational Method}
Our problems may be formulated as the systems of ordinary differential
equations
\begin{eqnarray}\label{eq:pol0}
& & Q_i(x)\frac{\ud x_i}{\ud t}=P_i(x,t),\quad x=(x_1,..., x_n),\\
& &\quad i=1,...,n, \quad
 \max_i  deg \ P_i=p, \quad \max_i deg \  Q_i=q \nonumber
\end{eqnarray}
with fixed initial conditions $x_i(0)$, where $P_i, Q_i$ are not more
than polynomial functions of dynamical variables $x_j$
and  have arbitrary dependence of time. Because of time dilation
we can consider  only next time interval: $0\leq t\leq 1$.
 Let us consider a set of
functions
\begin{eqnarray}
 \Phi_i(t)=x_i\frac{\ud}{\ud t}(Q_i y_i)+P_iy_i
\end{eqnarray}
and a set of functionals
\begin{eqnarray}
F_i(x)=\int_0^1\Phi_i (t)dt-Q_ix_iy_i\mid^1_0,
\end{eqnarray}
where $y_i(t) \ (y_i(0)=0)$ are dual (variational) variables.
It is obvious that the initial system  and the system
\begin{equation}\label{eq:veq}
F_i(x)=0
\end{equation}
are equivalent.
Of course, we consider such $Q_i(x)$ which do not lead to the singular
problem with $Q_i(x)$, when $t=0$ or $t=1$, i.e. $Q_i(x(0)), Q_i(x(1))\neq\infty$. 

In  part 5
we consider more general approach, which is based on possibility taking into
account underlying symplectic structure and on more useful and flexible
analytical approach, related to bilinear structure of initial functional.
Now we consider formal expansions for $x_i, y_i$:
\begin{eqnarray}\label{eq:pol1}
x_i(t)=x_i(0)+\sum_k\lambda_i^k\varphi_k(t)\quad
y_j(t)=\sum_r \eta_j^r\varphi_r(t),
\end{eqnarray}
where $\varphi_k(t)$ are useful basis functions of  some functional 
space ($L^2, L^p$, Sobolev, etc) corresponding to concrete 
problem and
 because of initial conditions we need only $\varphi_k(0)=0$.
\begin{equation}\label{eq:lambda}
\lambda=\{\lambda_i\}=\{\lambda^r_i\}=(\lambda_i^1, \lambda_i^2,...,\lambda_i^N), \quad
r=1,...,N, \quad i=1,...,n,
\end{equation}
 where the lower index i corresponds to 
expansion of dynamical variable with index i, i.e. $x_i$ and the upper index $r$
corresponds to the numbers of terms in the expansion of dynamical variables in the 
formal series.
Then we put (\ref{eq:pol1}) into the functional equations (\ref{eq:veq}) and as result
we have the following reduced algebraical system
of equations on the set of unknown coefficients $\lambda_i^k$ of
expansions (\ref{eq:pol1}):
\begin{eqnarray}\label{eq:pol2}
L(Q_{ij},\lambda,\alpha_I)=M(P_{ij},\lambda,\beta_J),
\end{eqnarray}
where operators L and M are algebraization of RHS and LHS of initial problem 
(\ref{eq:pol0}), where $\lambda$ (\ref{eq:lambda}) are unknowns of reduced system
of algebraical equations (RSAE)(\ref{eq:pol2}).

$Q_{ij}$ are coefficients (with possible time dependence) of LHS of initial 
system of differential equations (\ref{eq:pol0}) and as consequence are coefficients
of RSAE.

 $P_{ij}$ are coefficients (with possible time dependence) of RHS 
of initial system of differential equations (\ref{eq:pol0}) and as consequence 
are coefficients of RSAE.

$I=(i_1,...,i_{q+2}), \ J=(j_1,...,j_{p+1})$ are multiindexes, by which are 
labelled $\alpha_I$ and $\beta_I$ --- other coefficients of RSAE (\ref{eq:pol2}):
\begin{equation}\label{eq:beta}
\beta_J=\{\beta_{j_1...j_{p+1}}\}=\int\prod_{1\leq j_k\leq p+1}\varphi_{j_k},
\end{equation}
where p is the degree of polinomial operator P (\ref{eq:pol0})
\begin{equation}\label{eq:alpha}
\alpha_I=\{\alpha_{i_1}...\alpha_{i_{q+2}}\}=\sum_{i_1,...,i_{q+2}}\int
\varphi_{i_1}...\dot{\varphi_{i_s}}...\varphi_{i_{q+2}},
\end{equation}
where q is the degree of polynomial operator Q (\ref{eq:pol0}), 
$i_\ell=(1,...,q+2)$, $\dot{\varphi_{i_s}}=\ud\varphi_{i_s}/\ud t$.

Now, when we solve RSAE (\ref{eq:pol2}) and determine
 unknown coefficients from formal expansion (\ref{eq:pol1}) we therefore
obtain the solution of our initial problem.
It should be noted if we consider only truncated expansion (\ref{eq:pol1}) with N terms
then we have from (\ref{eq:pol2}) the system of $N\times n$ algebraical equations 
with degree $\ell=max\{p,q\}$
and
the degree of this algebraical system coincides
 with degree of initial differential system.
So, we have the solution of the initial nonlinear
(rational) problem  in the form
\begin{eqnarray}\label{eq:pol3}
x_i(t)=x_i(0)+\sum_{k=1}^N\lambda_i^k X_k(t),
\end{eqnarray}
where coefficients $\lambda_i^k$ are roots of the corresponding
reduced algebraical (polynomial) problem RSAE (\ref{eq:pol2}).
Consequently, we have a parametrization of solution of initial problem
by solution of reduced algebraical problem (\ref{eq:pol2}).
The first main problem is a problem of
 computations of coefficients $\alpha_I$ (\ref{eq:alpha}), $\beta_J$ 
(\ref{eq:beta}) of reduced algebraical
system.
As we will see, these problems may be explicitly solved in wavelet approach.

Next we consider the  construction  of explicit time
solution for our problem. The obtained solutions are given
in the form (\ref{eq:pol3}),
where
$X_k(t)$ are basis functions and
  $\lambda_k^i$ are roots of reduced
 system of equations.  In our first wavelet case $X_k(t)$
are obtained via multiresolution expansions and represented by
 compactly supported wavelets and $\lambda_k^i$ are the roots of
corresponding general polynomial  system (\ref{eq:pol2})  with coefficients, which
are given by CC or SSS  constructions.  According to the
        variational method   to  give the reduction from
differential to algebraical system of equations we need compute
the objects $\alpha_I$ and $\beta_J$.

\subsection{Wavelet Framework}
Our constructions are based on multiresolution approach. Because affine
group of translation and dilations is inside the approach, this
method resembles the action of a microscope. We have contribution to
final result from each scale of resolution from the whole
infinite scale of spaces. More exactly, the closed subspace
$V_j (j\in {\bf Z})$ corresponds to  level j of resolution, or to scale j.
We consider  a r-regular multiresolution analysis of $L^2 ({\bf R}^n)$
(of course, we may consider any different functional space)
which is a sequence of increasing closed subspaces $V_j$:
\begin{equation}
...V_{-2}\subset V_{-1}\subset V_0\subset V_{1}\subset V_{2}\subset ...
\end{equation}
satisfying the following properties:
\begin{eqnarray}
&&\displaystyle\bigcap_{j\in{\bf Z}}V_j=0,\quad
\overline{\displaystyle\bigcup_{j\in{\bf Z}}}V_j=L^2({\bf R}^n),\nonumber\\
&& f(x)\in V_j <=> f(2x)\in V_{j+1}, \nonumber\\
&& f(x)\in V_0 <=> f(x-k)\in V_0, \quad, \forall k\in {\bf Z}^n.
\end{eqnarray}
There exists a function $\varphi\in V_0$ such that
\{${\varphi_{0,k}(x)=
\varphi(x-k)}, k\in{\bf Z}^n$\} forms a Riesz basis for $V_0$.

The function $\varphi$ is regular and localized:
$\varphi$ is $C^{r-1}, \quad \varphi^{(r-1)}$ is almost
everywhere differentiable and for almost every $x\in {\bf R}^n$, for
every integer $\alpha\leq r$ and for all integer p there exists
constant $C_p$ such that
\begin{equation}
\mid\partial^\alpha \varphi(x)\mid \leq C_p(1+|x|)^{-p}
\end{equation}

Let
 $\varphi(x)$ be
a scaling function, $\psi(x)$ is a wavelet function and
$\varphi_i(x)=\varphi(x-i)$. Scaling relations that define
$\varphi,\psi$ are
\begin{eqnarray}
\varphi(x)&=&\sum\limits^{N-1}_{k=0}a_k\varphi(2x-k)=
\sum\limits^{N-1}_{k=0}a_k\varphi_k(2x),\\
\psi(x)&=&\sum\limits^{N-2}_{k=-1}(-1)^k a_{k+1}\varphi(2x+k).
\end{eqnarray}
Let  indices $\ell, j$
 represent translation and scaling, respectively and
\begin{equation}
\varphi_{jl}(x)=2^{j/2}\varphi(2^j x-\ell)
\end{equation}
then the set $\{\varphi_{j,k}\}, {k\in {\bf Z}^n}$ forms a Riesz basis for $V_j$.
The wavelet function $\psi $ is used to encode the details between
two successive levels of approximation.
Let $W_j$ be the orthonormal complement of $V_j$ with respect to $V_{j+1}$:
\begin{equation}
V_{j+1}=V_j\bigoplus W_j.
\end{equation}
Then just as $V_j$ is spanned by dilation and translations of the scaling function,
so are $W_j$ spanned by translations and dilation of the mother wavelet
$\psi_{jk}(x)$, where
\begin{equation}
\psi_{jk}(x)=2^{j/2}\psi(2^j x-k).
\end{equation}
All expansions which we used are based on the following properties:
\begin{eqnarray}
&&\{\psi_{jk}\}, \quad j,k\in {\bf Z}\quad
  \mbox{is a Hilbertian basis of } L^2({\bf R})\nonumber\\
&&\{\varphi_{jk}\}_{j\geq 0, k\in {\bf Z}} \quad\mbox{is an orthonormal
basis for} L^2({\bf R}),\nonumber\\
&& L^2({\bf R})=\overline{V_0\displaystyle\bigoplus^\infty_{j=0} W_j},\\
&& \mbox{or}\nonumber\\
&&\{\varphi_{0,k},\psi_{j,k}\}_{j\geq 0,k\in {\bf Z}} \quad\mbox{is
an orthonormal basis for}
 L^2({\bf R}).\nonumber
\end{eqnarray}

\subsection{ Wavelet Computations}
 Now we give construction for
computations of objects (17),(18) in the wavelet case.
We use compactly supported wavelet basis: orthonormal basis
for functions in $L^2({\bf R})$.

Let be  $ f : {\bf R}\longrightarrow {\bf C}$ and the wavelet
expansion is
\begin{eqnarray}
f(x)=\sum\limits_{\ell\in{\bf Z}}c_\ell\varphi_\ell(x)+
\sum\limits_{j=0}^\infty\sum\limits_{k\in{\bf
Z}}c_{jk}\psi_{jk}(x)
\end{eqnarray}

If in formulae (29) $c_{jk}=0$ for $j\geq J$, then $f(x)$ has an alternative
expansion in terms of dilated scaling functions only
$
f(x)=\sum\limits_{\ell\in {\bf Z}}c_{J\ell}\varphi_{J\ell}(x)
$.
This is a finite wavelet expansion, it can be written solely in
terms of translated scaling functions.
Also we have the shortest possible support: scaling function
$DN$ (where $N$ is even integer) will have support $[0,N-1]$ and
$N/2$ vanishing moments.
There exists $\lambda>0$ such that $DN$ has $\lambda N$
continuous derivatives; for small $N,\lambda\geq 0.55$.
To solve our second associated linear problem we need to
evaluate derivatives of $f(x)$ in terms of $\varphi(x)$.
Let be $
\varphi^n_\ell=\ud^n\varphi_\ell(x)/\ud x^n
$.
We consider computation of the wavelet - Galerkin integrals.
Let $f^d(x)$ be d-derivative of function
 $f(x)$, then we have
$
f^d(x)=\sum_\ell c_l\varphi_\ell^d(x)
$,
and values $\varphi_\ell^d(x)$ can be expanded in terms of
$\varphi(x)$
\begin{eqnarray}
\phi_\ell^d(x)&=&\sum\limits_m\lambda_m\varphi_m(x),\\
\lambda_m&=&\int\limits_{-\infty}^{\infty}\varphi_\ell^d(x)\varphi_m(x)\ud x,\nonumber
 \end{eqnarray}
where $\lambda_m$ are wavelet-Galerkin integrals.
The coefficients $\lambda_m$  are 2-term connection
coefficients. In general we need to find $(d_i\geq 0)$
\begin{eqnarray}
\Lambda^{d_1 d_2 ...d_n}_{\ell_1 \ell_2 ...\ell_n}=
 \int\limits_{-\infty}^{\infty}\prod\varphi^{d_i}_{\ell_i}(x)dx
\end{eqnarray}
For Riccati case we need to evaluate two and three
connection coefficients
\begin{eqnarray}
\Lambda_\ell^{d_1
d_2}=\int^\infty_{-\infty}\varphi^{d_1}(x)\varphi_\ell^{d_2}(x)dx,
\quad
\Lambda^{d_1 d_2
d_3}=\int\limits_{-\infty}^\infty\varphi^{d_1}(x)\varphi_
\ell^{d_2}(x)\varphi^{d_3}_m(x)dx
\end{eqnarray}
According to CC method [14] we use the next construction. When $N$  in
scaling equation is a finite even positive integer the function
$\varphi(x)$  has compact support contained in $[0,N-1]$.
For a fixed triple $(d_1,d_2,d_3)$ only some  $\Lambda_{\ell
 m}^{d_1 d_2 d_3}$ are nonzero: $2-N\leq \ell\leq N-2,\quad
2-N\leq m\leq N-2,\quad |\ell-m|\leq N-2$. There are
$M=3N^2-9N+7$ such pairs $(\ell,m)$. Let $\Lambda^{d_1 d_2 d_3}$
be an M-vector, whose components are numbers $\Lambda^{d_1 d_2
d_3}_{\ell m}$. Then we have the first reduced algebraical system
: $\Lambda$
satisfy the system of equations $(d=d_1+d_2+d_3)$
\begin{eqnarray}
A\Lambda^{d_1 d_2 d_3}=2^{1-d}\Lambda^{d_1 d_2 d_3},
\qquad
A_{\ell,m;q,r}=\sum\limits_p a_p a_{q-2\ell+p}a_{r-2m+p}
\end{eqnarray}
By moment equations we have created a system of $M+d+1$
equations in $M$ unknowns. It has rank $M$ and we can obtain
unique solution by combination of LU decomposition and QR
algorithm.
The second  reduced algebraical system gives us the 2-term connection
coefficients:
\begin{eqnarray}
A\Lambda^{d_1 d_2}=2^{1-d}\Lambda^{d_1 d_2},\quad d=d_1+d_2,\quad
A_{\ell,q}=\sum\limits_p a_p a_{q-2\ell+p}
\end{eqnarray}
For nonquadratic case we have analogously additional linear problems for
objects (31).
Solving these linear problems we obtain the coefficients of nonlinear
algebraical system (16) and after that we obtain the coefficients of wavelet
expansion (19). As a result we obtained the explicit time solution  of our
problem in the base of compactly supported wavelets.
We use for modelling D6, D8, D10 functions and programs RADAU and
DOPRI for testing.

In the following we consider extension of this approach to the case of periodic
boundary conditions, the case of presence of arbitrary variable coefficients
and more flexible biorthogonal wavelet approach.

\section{Variational Wavelet Approach for Periodic Trajectories}

We start with extension of our approach to the case
of periodic trajectories. The equations of motion corresponding
to our problems may be formulated as a particular case of
the general system of
ordinary differential equations
$
{dx_i}/{dt}=f_i(x_j,t)$, $  (i,j=1,...,n)$, $0\leq t\leq 1$,
where $f_i$ are not more
than rational functions of dynamical variables $x_j$
and  have arbitrary dependence of time but with periodic boundary conditions.
According to our variational approach we have the
solution in the following form
\begin{eqnarray}
x_i(t)=x_i(0)+\sum_k\lambda_i^k\varphi_k(t),\qquad x_i(0)=x_i(1),
\end{eqnarray}
where $\lambda_i^k$ are again the roots of reduced algebraical
systems of equations
with the same degree of nonlinearity and $\varphi_k(t)$
corresponds to useful type of wavelet bases (frames).
It should be noted that coefficients of reduced algebraical system
are the solutions of additional linear problem and
also
depend on particular type of wavelet construction and type of bases.

This linear problem is our second reduced algebraical problem. We need to find
in general situation objects
\begin{eqnarray}
\Lambda^{d_1 d_2 ...d_n}_{\ell_1 \ell_2 ...\ell_n}=
 \int\limits_{-\infty}^{\infty}\prod\varphi^{d_i}_{\ell_i}(x)\ud x,
\end{eqnarray}
but now in the case of periodic boundary conditions.
Now we consider the procedure of their
calculations in case of periodic boundary conditions
 in the base of periodic wavelet functions on
the interval [0,1] and corresponding expansion (35) inside our
variational approach. Periodization procedure
gives us
\begin{eqnarray}
\hat\varphi_{j,k}(x)&\equiv&\sum_{\ell\in Z}\varphi_{j,k}(x-\ell)\\
\hat\psi_{j,k}(x)&=&\sum_{\ell\in Z}\psi_{j,k}(x-\ell)\nonumber
\end{eqnarray}
So, $\hat\varphi, \hat\psi$ are periodic functions on the interval
 [0,1]. Because $\varphi_{j,k}=\varphi_{j,k'}$ if $k=k'\mathrm{mod}(2^j)$, we
may consider only $0\leq k\leq 2^j$ and as  consequence our
multiresolution has the form
$\displaystyle\bigcup_{j\geq 0} \hat V_j=L^2[0,1]$ with
$\hat V_j= \mathrm{span} \{\hat\varphi_{j,k}\}^{2j-1}_{k=0}$ [15].
Integration by parts and periodicity gives  useful relations between
objects (36) in particular quadratic case $(d=d_1+d_2)$:
\begin{eqnarray}
\Lambda^{d_1,d_2}_{k_1,k_2}=(-1)^{d_1}\Lambda^{0,d_2+d_1}_{k_1,k_2},
\Lambda^{0,d}_{k_1,k_2}=\Lambda^{0,d}_{0,k_2-k_1}\equiv
\Lambda^d_{k_2-k_1}
\end{eqnarray}
So, any 2-tuple can be represent by $\Lambda^d_k$.
Then our second additional linear problem is reduced to the eigenvalue
problem for
$\{\Lambda^d_k\}_{0\leq k\le 2^j}$ by creating a system of $2^j$
homogeneous relations in $\Lambda^d_k$ and inhomogeneous equations.
So, if we have dilation equation in the form
$\varphi(x)=\sqrt{2}\sum_{k\in Z}h_k\varphi(2x-k)$,
then we have the following homogeneous relations
\begin{equation}
\Lambda^d_k=2^d\sum_{m=0}^{N-1}\sum_{\ell=0}^{N-1}h_m h_\ell
\Lambda^d_{\ell+2k-m},
\end{equation}
or in such form
$A\lambda^d=2^d\lambda^d$, where $\lambda^d=\{\Lambda^d_k\}_
{0\leq k\le 2^j}$.
Inhomogeneous equations are:
\begin{equation}
\sum_{\ell}M_\ell^d\Lambda^d_\ell=d!2^{-j/2},
\end{equation}
 where objects
$M_\ell^d(|\ell|\leq N-2)$ can be computed by recursive procedure
\begin{equation}
M_\ell^d=2^{-j(2d+1)/2}\tilde{M_\ell^d}, \quad\tilde{M_\ell^k}=
<x^k,\varphi_{0,\ell}>=\sum_{j=0}^k {k\choose j} n^{k-j}M_0^j,\quad
\tilde{M_0^\ell}=1.
\end{equation}
 So, we reduced our last problem to standard
linear algebraical problem. Then we use the same methods as in part 3.3.
As a result we obtained
for closed trajectories of orbital dynamics  
the explicit time solution (35) in the base of periodized wavelets (37).

\section{Variational Approach in Biorthogonal\\ Wavelet Bases}

Now we consider further generalization of our variational wavelet approach.
Because integrand of variational functionals is represented
by bilinear form (scalar product) it seems more reasonable to
consider wavelet constructions [16] which take into account all advantages of
this structure.

The action functional for loops in the phase space is 
\begin{equation}
F(\gamma)=\displaystyle\int_\gamma pdq-\int_0^1H(t,\gamma(t))dt
\end{equation}
The critical points of $F$ are those loops $\gamma$, which solve
the Hamiltonian equations associated with the Hamiltonian $H$
and hence are periodic orbits.

Let us consider the loop space $\Omega=C^\infty(S^1, R^{2n})$,
where $S^1=R/{\bf Z}$, of smooth loops in $R^{2n}$.
Let us define a function $\Phi: \Omega\to R $ by setting
\begin{equation}
\Phi(x)=\displaystyle\int_0^1\frac{1}{2}<-J\dot x, x>dt-
\int_0^1 H(x(t))dt, \quad x\in\Omega
\end{equation}
The critical points of $\Phi$ are the periodic solutions of $\dot x=X_H(x)$.
Computing the derivative at $x\in\Omega$ in the direction of $y\in\Omega$,
we find
\begin{eqnarray}
\Phi'(x)(y)=\frac{d}{d\epsilon}\Phi(x+\epsilon y)\vert_{\epsilon=0}
=
\displaystyle\int_0^1<-J\dot x-\bigtriangledown H(x),y>dt
\end{eqnarray}
Consequently, $\Phi'(x)(y)=0$ for all $y\in\Omega$ iff the loop $x$ satisfies
the equation
\begin{equation}
-J\dot x(t)-\bigtriangledown H(x(t))=0,
\end{equation}
i.e. $x(t)$ is a solution of the Hamiltonian equations, which also satisfies
$x(0)=x(1)$, i.e. periodic of period 1.

But now we
need to take into account underlying bilinear structure via wavelets.

We started with two hierarchical sequences of approximations spaces [16]:
\begin{eqnarray}
\dots V_{-2}\subset V_{-1}\subset V_{0}\subset V_{1}\subset V_{2}\dots,\qquad
\dots \widetilde{V}_{-2}\subset\widetilde{V}_{-1}\subset
\widetilde{V}_{0}\subset\widetilde{V}_{1}\subset\widetilde{V}_{2}\dots, \nonumber
\end{eqnarray}
and as usually,
$W_0$ is complement to $V_0$ in $V_1$, but now not necessarily orthogonal
complement.
New orthogonality conditions have now the following form:
\begin{equation}
\widetilde {W}_{0}\perp V_0,\qquad  W_{0}\perp\widetilde{V}_{0},\qquad
V_j\perp\widetilde{W}_j, \qquad \widetilde{V}_j\perp W_j
\end{equation}
translates of $\psi$ $\mathrm{span}$ $ W_0$,
translates of $\tilde\psi \quad \mathrm{span} \quad\widetilde{W}_0$.
Biorthogonality conditions are
\begin{equation}
<\psi_{jk},\tilde{\psi}_{j'k'}>=
\int^\infty_{-\infty}\psi_{jk}(x)\tilde\psi_{j'k'}(x)\ud x=
\delta_{kk'}\delta_{jj'},
\end{equation}
 where
$\psi_{jk}(x)=2^{j/2}\psi(2^jx-k)$.
Functions $\varphi(x), \tilde\varphi(x-k)$ form  dual pair:
\begin{equation}
<\varphi(x-k),\tilde\varphi(x-\ell)>=\delta_{kl},\quad
 <\varphi(x-k),\tilde\psi(x-\ell)>=0\quad  \mbox{for}\quad \forall k,
\ \forall\ell.
\end{equation}
Functions $\varphi, \tilde\varphi$ generate a multiresolution analysis.
$\varphi(x-k)$, $\psi(x-k)$ are synthesis functions,
$\tilde\varphi(x-\ell)$, $\tilde\psi(x-\ell)$ are analysis functions.
Synthesis functions are biorthogonal to analysis functions. Scaling spaces
are orthogonal to dual wavelet spaces.
Two multiresolutions are intertwining
$
V_j+W_j=V_{j+1}, \quad \widetilde V_j+ \widetilde W_j = \widetilde V_{j+1}
$.
These are direct sums but not orthogonal sums.

So, our representation for solution has now the form
\begin{equation}
f(t)=\sum_{j,k}\tilde b_{jk}\psi_{jk}(t),
\end{equation}
where synthesis wavelets are used to synthesize the function. But
$\tilde b_{jk}$ come from inner products with analysis wavelets.
Biorthogonality yields
\begin{equation}
\tilde b_{\ell m}=\int f(t)\tilde{\psi}_{\ell m}(t) \ud t.
\end{equation}
So, now we can introduce this more complicated construction into
our variational approach. We have modification only on the level of
computing coefficients of reduced nonlinear algebraical system.
This new construction is more flexible.
Biorthogonal point of view is more stable under the action of large
class of operators while orthogonal (one scale for multiresolution)
is fragile, all computations are much more simpler and we accelerate
the rate of convergence. In all types of (Hamiltonian) calculation,
which are based on some bilinear structures (symplectic or
Poissonian structures, bilinear form of integrand in variational
integral) this framework leads to greater success.

\section{Variable Coefficients}
In the case when we have situation when our problem is described by a system of
nonlinear (rational) differential equations, we need to consider
extension of our previous approach which can take into  account
any type of variable coefficients (periodic, regular or singular).
We can produce such approach if we add in our construction additional
refinement equation, which encoded all information about variable
coefficients [17].
According to our variational approach we need to compute integrals of
the form
\begin{equation}\label{eq:var1}
\int_Db_{ij}(t)(\varphi_1)^{d_1}(2^m t-k_1)(\varphi_2)^{d_2}
(2^m t-k_2)\ud x,
\end{equation}
where now $b_{ij}(t)$ are arbitrary functions of time, where trial
functions $\varphi_1,\varphi_2$ satisfy a refinement equations:
\begin{equation}
\varphi_i(t)=\sum_{k\in{\bf Z}}a_{ik}\varphi_i(2t-k)
\end{equation}
If we consider all computations in the class of compactly supported wavelets
then only a finite number of coefficients do not vanish. To approximate
the non-constant coefficients, we need choose a different refinable function
$\varphi_3$ along with some local approximation scheme
\begin{equation}
(B_\ell f)(x):=\sum_{\alpha\in{\bf Z}}F_{\ell,k}(f)\varphi_3(2^\ell t-k),
\end{equation}
where $F_{\ell,k}$ are suitable functionals supported in a small neighborhood
of $2^{-\ell}k$ and then replace $b_{ij}$ in (\ref{eq:var1}) by
$B_\ell b_{ij}(t)$. In particular case one can take a characteristic function
and can thus approximate non-smooth coefficients locally. To guarantee
sufficient accuracy of the resulting approximation to (\ref{eq:var1})
it is important to have the flexibility of choosing $\varphi_3$ different
from $\varphi_1, \varphi_2$. In the case when D is some domain, we
can write
\begin{equation}
b_{ij}(t)\mid_D=\sum_{0\leq k\leq 2^\ell}b_{ij}(t)\chi_D(2^\ell t-k),
\end{equation}
where $\chi_D$ is characteristic function of D. So, if we take
$\varphi_4=\chi_D$, which is again a refinable function, then the problem of
computation of (\ref{eq:var1}) is reduced to the problem of calculation of
integral
\begin{eqnarray}
&&H(k_1,k_2,k_3,k_4)=H(k)=\\
&&\int_{{\bf R}^s}\varphi_4(2^j t-k_1)\varphi_3(2^\ell t-k_2)
\varphi_1^{d_1}(2^r t-k_3)
\varphi_2^{d_2}(2^st-k_4)\ud x\nonumber
\end{eqnarray}
The key point is that these integrals also satisfy some sort of refinement
equation [17]:
\begin{equation}
2^{-|\mu|}H(k)=\sum_{\ell\in{\bf Z}}b_{2k-\ell}H(\ell),\qquad \mu=d_1+d_2.
\end{equation}

This equation can be interpreted as the problem of computing an eigenvector.
Thus, we reduced the problem of extension of our method to the case of
variable coefficients to the same standard algebraical problem as in
the preceding sections. So, the general scheme is the same one and we
have only one more additional
linear algebraic problem by which we in the same way can parameterize the
solutions of corresponding problem.

\section{Numerical Calculations}

In this part we consider numerical illustrations of previous analytical
approach. Our numerical calculations are based on compactly supported
Daubechies wavelets and related wavelet families.

On Fig.~2 we present according to formulae (2) contributions 
to approximation of our dynamical evolution (top row on the Fig.~3) starting from
the coarse approximation, corresponding to scale $2^0$ (bottom row)
to the finest one corresponding to the scales from $2^1$ to  $2^5$
or from slow to fast components (5 frequencies) as details for approximation.
Then on Fig.~3, from bottom to top, we demonstrate the summation
of contributions from corresponding levels of resolution given on
Fig.~2 and as result we restore via 5 scales (frequencies) approximation
our dynamical process(top row on Fig.~3 ).
In this particular model case we considered for
approximation simple two frequencies harmonic
process.
But the same situation we have on the Fig.~5 and Fig.~6 in case when we added
to previous 2-frequencies harmonic process the noise as perturbation.
Again, our dynamical process under investigation (top row of Fig.~6) is recovered
via 5 scales contributions (Fig.~5) to approximations (Fig.~6).
The same decomposition/approximation we produce also on the level of
power spectral density in the process without noise (Fig.~4)
and with noise (Fig.~7).
On Fig.~8 we demonstrate the family of localized contributions
to beam motion, which we also may consider for such type of approximation.

It should be noted that complexity of such algorithms are minimal regarding
other possible.
Of course, we may use different multiresolution analysis schemes, which
are based on different families of generating wavelets and apply
such schemes of numerical--analytical calculations to any dynamical
process which may be described by systems of ordinary/partial differential
equations with rational nonlinearities [13].

\begin{figure}
\centering
\epsfig{file=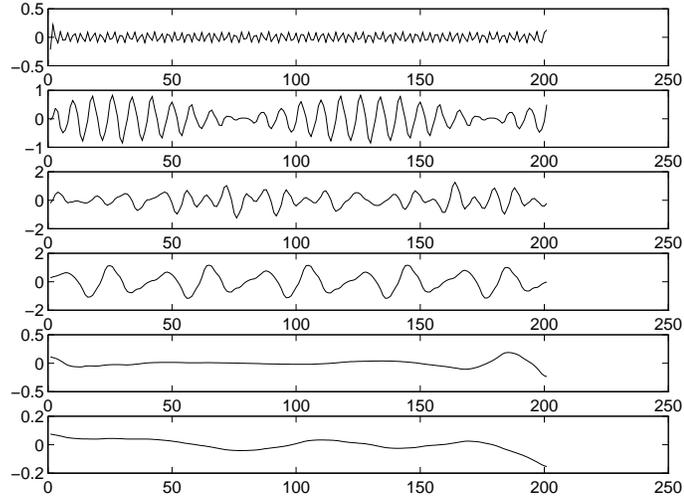, width=90mm}
\caption{Contributions to approximation: from scale $2^1$ to $2^5$ (without noise).}
\end{figure}
\begin{figure}
\centering
\epsfig{file=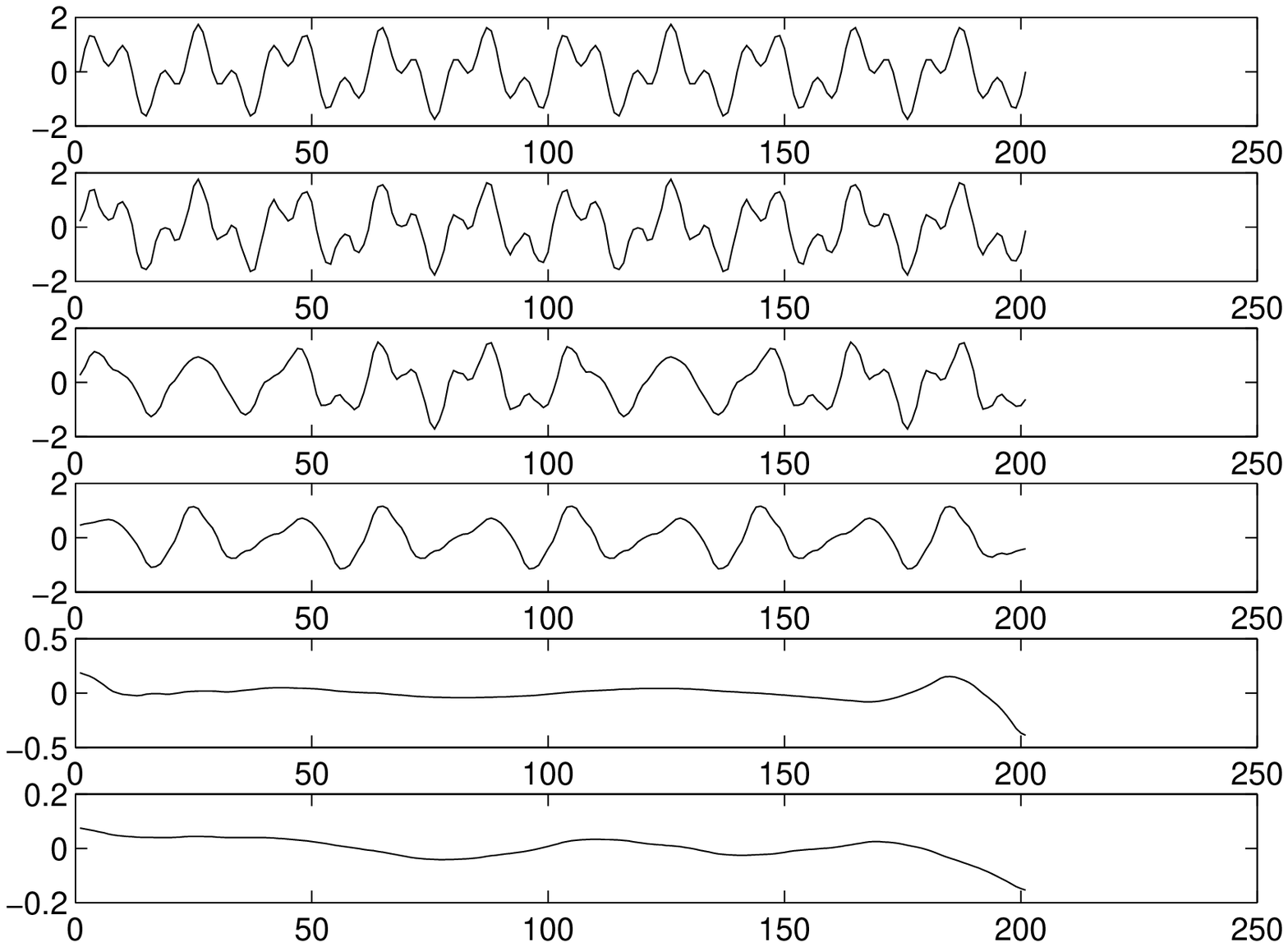, width=90mm}
\caption{Approximations: from scale $2^1$ to $2^5$ (without noise).}
\end{figure}
\begin{figure}
\centering
\epsfig{file=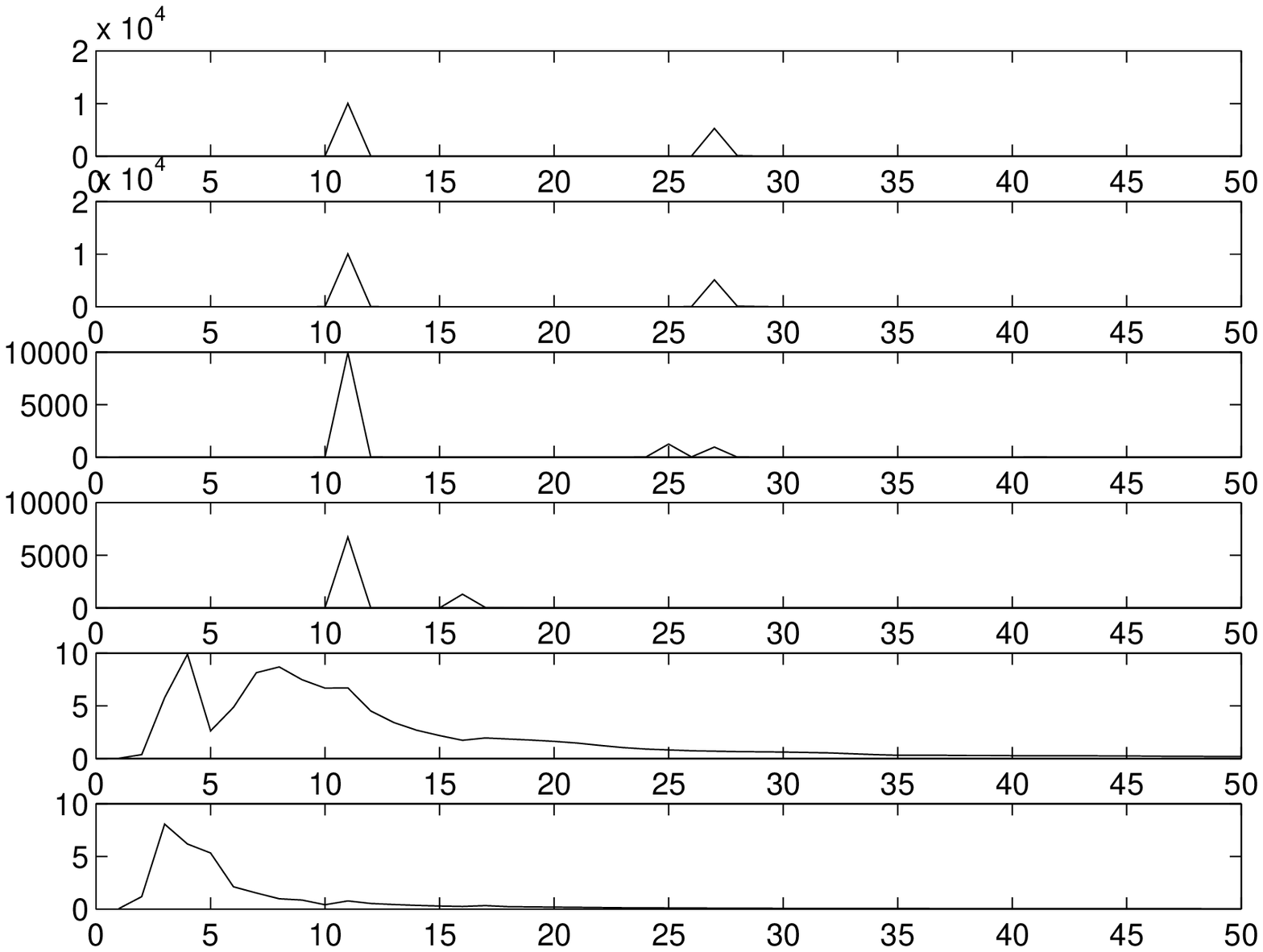, width=90mm}
\caption{Power spectral density:  from scale $2^1$ to $2^5$ (without noise)}
\end{figure}
\begin{figure}
\centering
\epsfig{file=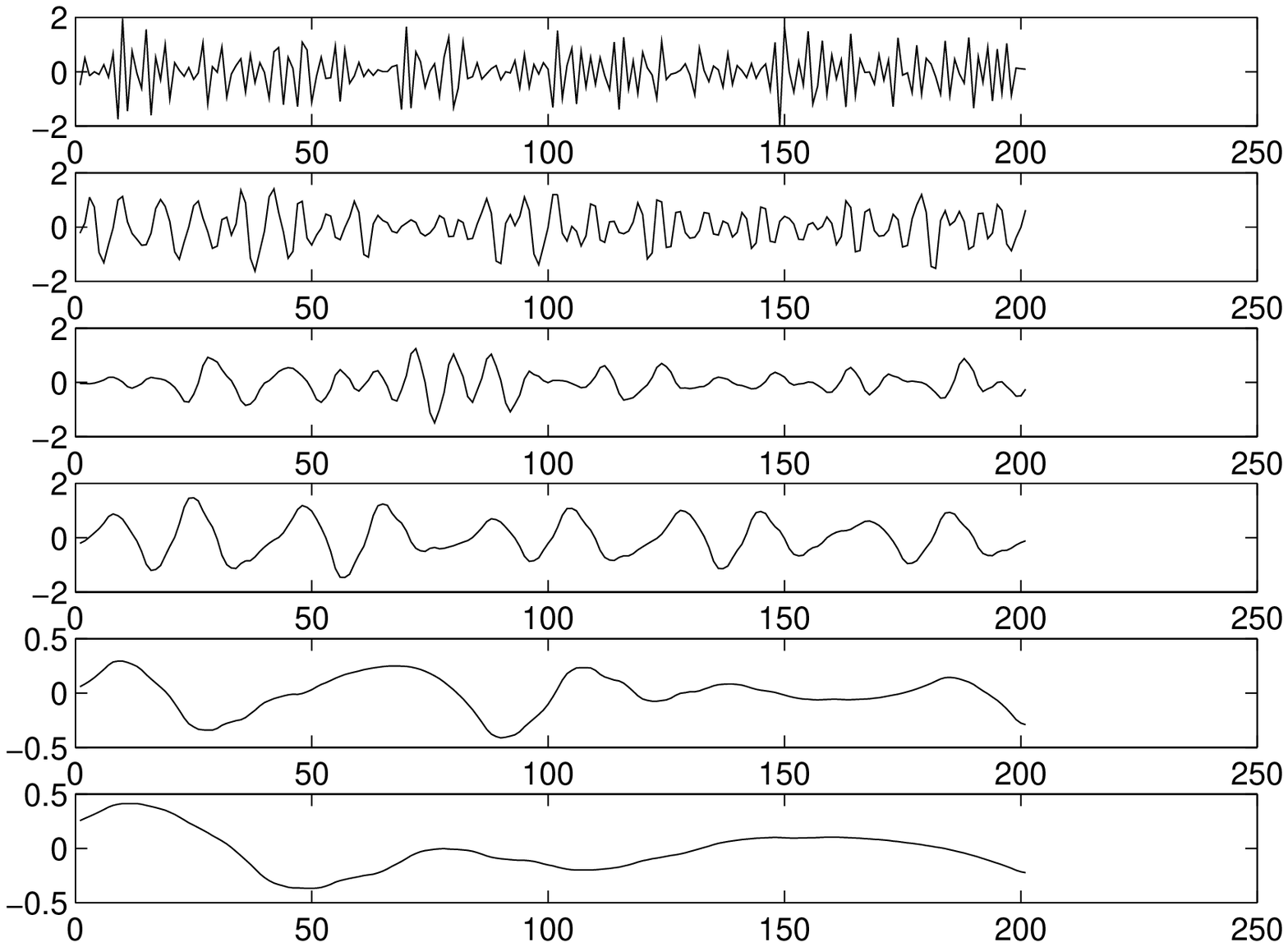, width=90mm}
\caption{Contributions to approximation: from scale $2^1$ to $2^5$ (with noise).}
\end{figure}
\begin{figure}
\centering
\epsfig{file=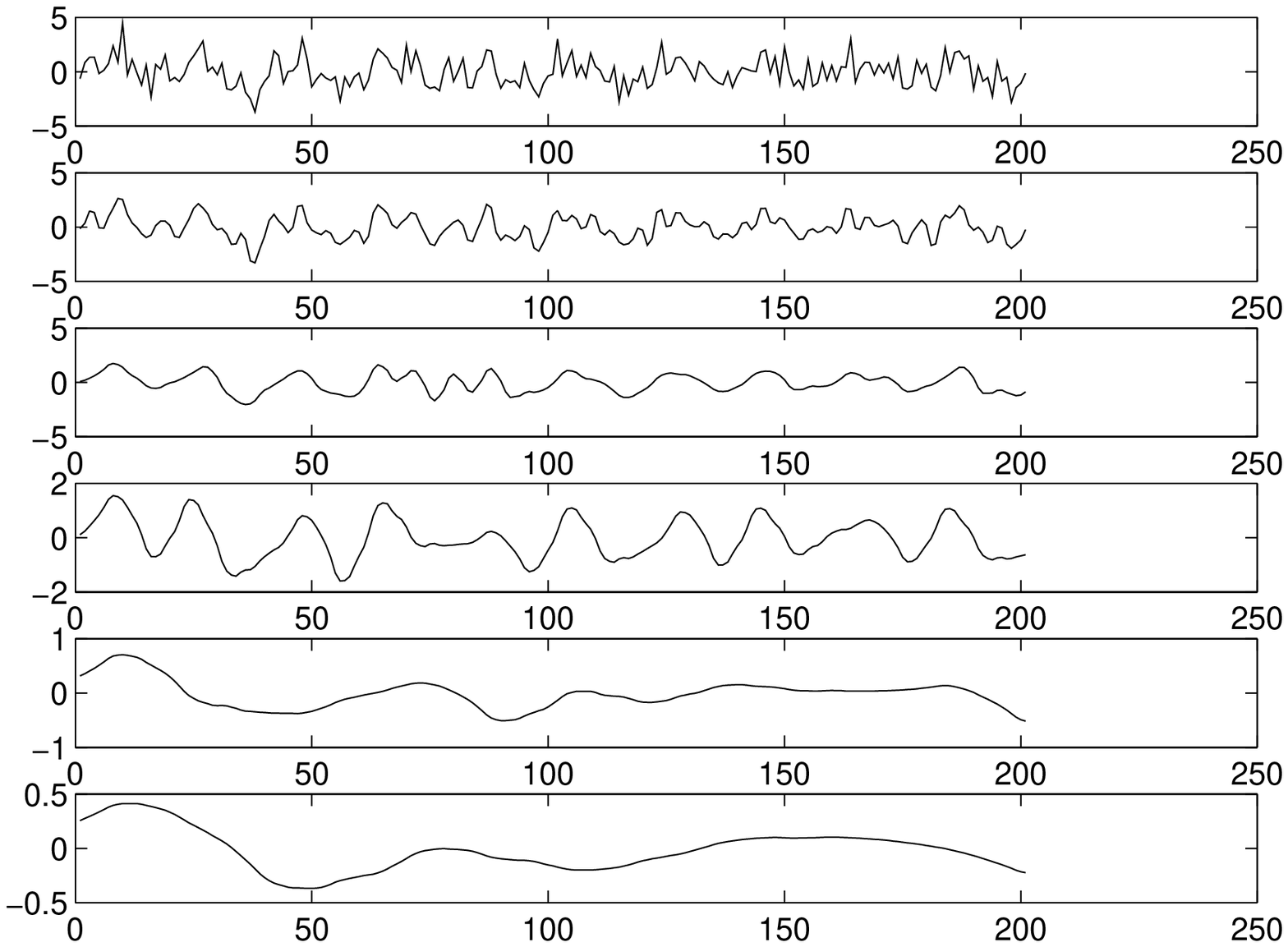, width=90mm}
\caption{Approximations: from scale $2^1$ to $2^5$ (with noise). }
\end{figure}
\begin{figure}
\centering
\epsfig{file=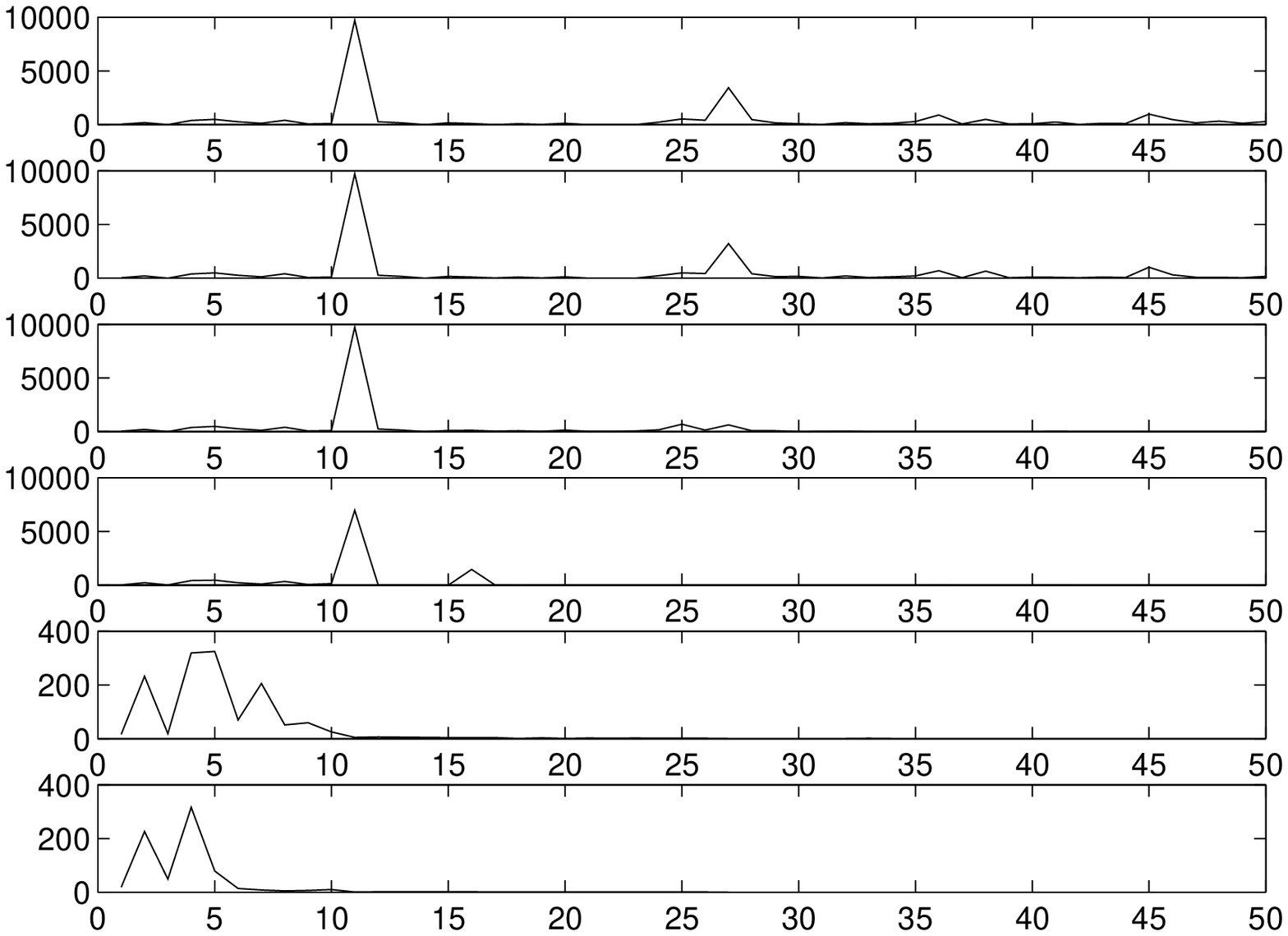, width=90mm}
\caption{Power spectral density: from scale $2^1$ to $2^5$ (with noise)}
\end{figure}
\begin{figure}[ht]
\centering
\epsfig{file=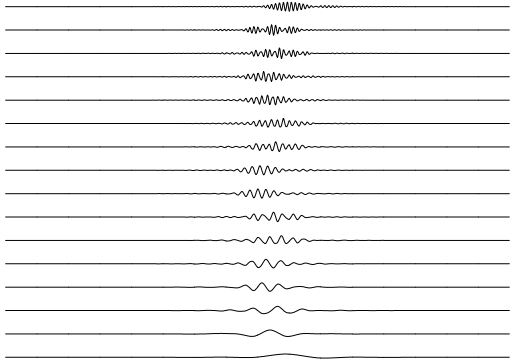, width=90mm, bb=0 200 599 590, clip}
\caption{Localized contributions to beam motion.}
\end{figure}

\newpage

\section*{Acknowledgments}
We would like to thank Professor
James B. Rosenzweig and Mrs. Melinda Laraneta for
nice hospitality, help, support and discussions before and during Workshop
and all participants for interesting discussions.

 \end{document}